\documentclass[12pt,preprint]{aastex} 

\shorttitle{The genesis of morphologies in radio sources} 
\shortauthors{Saripalli and Subrahmanyan} 
 
\begin{document} 
 
\title{The genesis of morphologies in extended radio sources: X-shapes,
  off-axis distortions and giant radio sources.} 
 
\author{Lakshmi Saripalli \& Ravi Subrahmanyan} 
\affil{Raman Research Institute, C. V. Raman Avenue, Sadashivanagar, Bangalore 560080,  
India} 
\email{lsaripal@rri.res.in (LS), rsubrahm@rri.res.in (RS)} 
 
\begin{abstract} 

We examine relationships between the morphology in double radio
sources and the radio-optical position angle offset---the
relative orientation of the radio axis with respect to the major axis of the
host galaxy. The study was done for a representative sample of radio
sources: the nearby (redshift $z < 0.5$) 3CRR sources, and separately for
samples of giant radio sources and X-shaped radio sources.  We find that
radio morphological features have a dependence on the radio-optical position
angle offset and on whether the source is a major- or minor-axis source. The
evidence indicates an anisotropic galaxy environment, related to the
ellipticity of the host galaxy, that causes the source linear size
evolution, strength of backflow in the radio lobes, off-axis lobe distortions
and the formation of wings and X-shaped radio sources to depend on the
radio-optical position angle offset. We identify a class of
X-shaped radio sources, which are either edge-darkened or lacking hotspots,
and appear to have inner doubles suggesting a restarting of activity. We
suggest a common formation mechanism, requiring backflows, for these
apparently FR-I X-shaped radio sources as well as the edge-brightened X-shaped
sources. 

\end{abstract}


\section{Introduction} 
 
The extended radio morphologies in extragalactic 
radio sources manifest in a variety of structures. However, a  
basic double-lobed structure is invariably present and is accepted as fundamental to   
the nature of these radio sources. Among the various departures from this basic   
structure, radio galaxies with X-shaped morphologies have staked 
claim to the bizarre. Apart from the usual double lobes, prototypical X-shaped
radio sources have a second pair of lobes---sometimes larger than the main
lobe pair---sharing the same radio core but oriented
along a vastly different axis. Some of the most impressive examples of this 
class of X-shaped radio galaxies are 4C12.03, 3C223.1, 3C315 and 3C403
\citep{lea91, bla92, lea84}.
 
Extended powerful radio sources typically has hotspots at the ends of the lobe
pair, and diffuse emission extending from the hotspots
towards the core, often forming a continuous bridge of relatively lower
surface brightness. \citet{lea84} studied and classified  
distortions to normal bridge structures: they were found to be quite common, occurring in as  
many as $60\%$ of sources, and the distortions occurred
predominantly at the centres with either one or both lobes deflecting off the
radio axis. These lobe deviations at the  
centres were suggested to be a consequence of interaction between the strong backflows and  
a gaseous halo associated with the host galaxy. Inversion symmetry in the
central distortions involved backflow  
material encountering an asymmetrically distributed galaxy halo.   
The large deviations observed in X-shaped radio sources were suggested as 
resulting from a deflection of backflowing material into a cavity created by past 
activity along a very different direction.  
 
Recent phenomenological studies of the extended emission associated with
active galactic nuclei (AGN) have provided additional clues to the genesis of
the radio morphologies. \citet{cap02} presented  
evidence for a preference for X-shaped radio galaxies to be hosted by 
galaxies with relatively high ellipticities and for the lobe deflections to be oriented 
close to the host minor axes. Based on their findings a model for formation of
X-shaped structures was proferred wherein backflows occurring closer to host
major axes responded to the high cocoon pressure by escaping along the
direction of maximum pressure gradient, which  
is along the host galaxy minor axis. While the latter finding of the `wings' being  
preferentially along host minor axes has been subsequently confirmed,
observational support for the stringent 
requirement for high host ellipticities has not been forthcoming \citep{spr07}.
Supporting this model, X-ray emission corresponding to
a thermal gaseous halo has been detected in the host of an X-shaped radio
source 3C403: the highly elliptic X-ray contours of the halo follow the
elliptic galaxy isophotes and 
appear tangential to the sharp bend in the radio contours of the backflow,
suggesting a deflection interaction \citep{kra05}. 

\citet{sar08} presented an interesting case of a giant, FR-I type \citep{fan74} 
X-shaped radio galaxy that obeys the same  
empirical relations found by \citet{cap02} for FR-II X-shaped sources. The 
discovery of its restarted nature via the finding of an inner, edge-brightened double source 
sharing the same core and radio axis suggests a model consistent with the backflow 
scenario. 

In an attempt to account for the Giant Metrewave Radio
Telescope (GMRT) observations of anomalous spectral indices in the twin lobes, 
\citet{lal07} proposed that the X-shapes originated in 
two closely spaced AGNs where each was associated with a pair of lobes
whose radio axes are at a large angle with respect to each other.

On the other hand, considering the observational evidence for episodic activity  
associated with the AGNs in several radio sources
\citep{roe94,sub96,sch00,sar02,sar03,sai06}, the obvious
question is whether X-shaped radio sources are a result of a restarting of
beams following a flip in the ejection axis \citep{lea84,zie01,mer02,liu04}.  This
alternative model---the  
binary black-hole merger model---suggests the eventual merger of two black
holes, one of which is responsible for the radio jets, as causing 
a rapid flip in the spin axis resulting in a new double radio 
source along a very different axis. 
This is the preferred model for the multi-frequency
observations of two X-shaped radio galaxies presented by   
\citet{den02},  and for the double-peaked low-ionization emission lines
observed by \citet{zha07} in an X-shaped source. 
The existence of wings that are larger than the main source extent has been
viewed as an argument against the backflow model \citep{lea92}; such
structures are accounted for in the black-hole merger
model as relic emission from an independent past activity epoch, which may
have been ongoing for longer time compared to the current actvity. However, as
has been pointed out by \citet{kra05}, it is possible, within the backflow model, for a very
asymmetrical environment and fast backflows to result in wings advancing
faster than the main lobe.  


We take up the issue of the formation of 
X-shaped structures in this paper, expanding the discussion to the formation
of off-axis deformations that are less spectacular, and including the question
of formation of giant radio galaxies.  We first discuss the place of FR-I
X-shaped sources in Section~2. We then explore the
relationships between the radio axis and host galaxy major axis for several
samples, in Sections~3 to 6, representing different radio morphological
structures. We find evidence that a variety of radio morphological features
might be a result of 
interactions between the jets, backflows, and an asymmetric environment, and
the discussion in Section~7 consolidates the viewpoint that the radio
morphologies are related to whether the source is of the minor- or major-axis
class, as defined by the relationship between the radio axis and optical axes.  
 
At the outset we define an X-shaped source to be one that has extensions on
either side of the   
core in addition to the main lobes, with at least one extension having a total
extent more than $50\%$ of the main lobe; such extensions are referred to
herein as `wings'. Extensions that are less than $50\%$ of the   
corresponding lobes are called `mini-wings'.  
We adopt a flat cosmology with Hubble constant $H_{0}$ = 71~km~sec$^{-1}$~Mpc$^{-1}$ and matter  
density parameter $\Omega_{m}$ = 0.27. 
 
\section{Edge-darkened and hotspot-less X-shaped radio galaxies} 
 
X-shaped structures are almost always found in edge-brightened radio sources. 
This aspect has been an important clue to the cause of the X-shaped structures,  
and has been used to argue for or against models proposed for the formation of
the X-structures.    

In models that require backflows and invoke deflection of backflowing lobe material for the  
creation of the prominent wings, edge-brightened radio structures---indicative of powerful    
jets and compact hotspots---are necessary. In this class of models  
FR-I sources would not be expected to show X-shaped structures.  On the other
hand, the lack of   
FR-I X-shaped sources has been viewed as a puzzle or attributed to negligible 
interaction between the black hole binary and radiatively inefficient accretion
disks in FR-Is \citep{den02,liu04}. It has also been used as an argument against
models in which the   
wings are relics of past activity, which was along a different position angle
\citep{cap02}. 
 
This belief that X-shaped radio sources are exclusively edge-brightened and hot-spot radio  
galaxies---owing to the lack of observations of X structures in edge-darkened radio
sources---has been dented by the recent discovery   
and imaging of the clearly FR-I X-shaped radio galaxy B2014$-$558 \citep{sar08}.   
Subsequently, we have carefully examined X-shaped radio galaxies in the literature and have  
identified several more examples, which have either edge-darkened lobes or
edge-brightened lobes   
that lack hotspots.  Examples of FR-I type edge-darkened X-shaped sources, or
X-shaped sources   
lacking hotspots at the ends, are considered a clear argument against models
that require backflows for the   
formation of the secondary radio axis. Since the   
sources we have identified have no hotspots at the ends of their lobes,
backflows are not expected.   
Therefore, it would appear that their secondary radio axes could not have formed via the  
deflection of backflows, suggesting that X-shaped structures---at least in such  
cases---might form via flips in the radio ejection axis. 
 
 Below, we list and provide brief notes on the X-shaped radio galaxies that we
 have identified to   
have edge-darkened lobes, or lacking hotspots at their lobe ends.  All of
these have secondary   
radio structures or `wings' that are at least 50\% of the extent of the main radio axis. 
 
 (1) B2014$-$558:~A study of the radio properties of this unusual
 FR-I X-shaped radio galaxy has been presented by \citet{sar08}. To our knowledge,   
it is the first clearly FR-I type X-shaped radio galaxy to be recognized as
such and studied in detail. The source is atypical in more than one
aspect.  It is the only known giant radio source with the
classic X-structure.  As discovered by \citet{sar08}, B2014$-$558 is also 
a double-double radio
source and has an inner double at the centre, aligned with the 1.5-Mpc outer
lobes, and nearly two orders of magnitude smaller in linear extent. 
 
 (2) 4C12.03:~Although the radio source appears to have an edge-brightened
 morphology, the lobes do not   
possess compact hotspots at the ends \citep{lea91}. 4C12.03 has one of the
most prominent set of wings among   
known X-shaped radio sources: the wings span an extent that is larger than
that of the main radio axis.    
Once again, the radio source is of the double-double category and has an
embedded inner double at the   
centre, with extent about 30~kpc, that is observed to be collinear with the main lobes.  
 
 (3) 3C315:~This is a well studied X-shaped radio source \citep{lea84} with
 clearly edge-darkened lobes,   
although the northern lobe has a warm-spot located somewhat away from the
end. There are no compact   
hotspots in either of the two main lobes.  Apart from the relaxed appearance
for the lobes along the   
main axis and the lack of hot spots, which are unusual in X-shaped radio
sources, 3C315 appears to   
possess a relatively small, inner edge-brightened double radio source at the centre as in 
the previous two cases presented above. The inner-double is about 8~kpc in
extent and is closely aligned with the axis of the main source 
(3CRR Atlas\footnotemark[1], eds. J. P. Leahy, A. H. Bridle, R. G. Strom). 

\footnotetext[1]{http://www.jb.man.ac.uk/atlas/}
 
 (4) J0116$-$473:~\citet{sar02} presented a detailed study of this radio
 source, pointing out that it has   
a double-double structure and is a spectacular example of recurrent nuclear
activity in which new   
edge-brightened inner lobes have been created embedded within outer
edge-brightened relic lobes of   
past activity; the outer lobes lack hotspots. Additionally, they drew
attention to a peculiar distinct   
bar-like feature that is associated with the inner part of the southern lobe
and offset from the core,   
and is oriented almost orthogonal to the radio source axis.  If we consider
this bar as a `wing'   
then J0116$-$473 is another example of a restarting X-shaped radio galaxy that
lacks hotspots at the lobe ends. 
 
 (5) B1531+104:~This source has been observed by \citet{owe97} and is a radio
 source in an Abell cluster.    
The source has wings that are asymmetric in extent: the eastern wing is less
than half the extent of the  
western wing.  There are no hotspots in both of the main pair of lobes. As in
previous examples listed   
above, B1531+104 is observed to have an inner double structure that
constitutes all of the bright structure in the source and is aligned with the
main axis as defined by the main pair of lobes.  
 
 (6) B1207+722:~This source is also a member of an Abell cluster and was
 imaged by \citet{owe97}.  The   
observations clearly show a pair of wings centered on the core. The two main
lobes have their brightest   
regions well recessed from the lobe ends; this attribute is more manifest for
the northern lobe.   
 
 (7) NGC~326:~Radio images of this source have been presented by
 \citet{mur01}, and it has been noted that both of the pair of main   
lobes lack compact hotspots at their ends. The wings in NGC~326 are among the
most spectacular of such   
structures: they are more than twice the main source in extent. The host is
part of a dumb-bell galaxy pair.  
 
(8) 3C76.1:~This source has a clear edge-darkened morphology although a pair of warm-spots are 
present at the lobe ends \citep{mac83}. The source lacks the classic bright pair of jets;
instead, it has a  
nested pair of emission peaks on either side of the core. The S-shaped
morphology of the main axis  
suggests that the jet may have precessed over a small angle. The source has a
wide wing that is  
centred on the host galaxy.  
 
For each of the above eight sources we have measured the sky angle between
the radio axis, as defined by   
the main pair of radio lobes, and the major axis of the host galaxy.  This has
been relatively error free   
in six cases (B2014$-$558, 4C12.03, 3C315, J0116$-$473, B1207+722 and 3C76.1)
in which the host galaxy images are clearly non-circular.  It is 
significant that in all   
of these six cases the radio axis is oriented close to the major axis of the
host galaxy, and the   
relative angle between the radio axis and major axis of the host is within
$45\degr$.  The sources   
have been selected based on the radio morphology, with no bias related to the
host optical properties   
and, therefore, we believe that our result is a finding of a genuine propensity.   
 
It is remarkable that all of the six X-shaped radio sources that lack
hotspots at the ends---with some,   
additionally, having edge-darkened lobes---show the same alignment between the
main radio axis and the   
major axis of the host galaxy as was reported by \citet{cap02} for FR-II
X-shaped radio sources.  Moreover, in all six cases with clearly non-circular
hosts, the wing-lobe pairs lie on the same side of their host galaxy major
axis, suggesting that the wings are a result of interactions of backflow from
the corresponding lobe and an asymmetric environment associated with the host galaxy.
Although it may seem that the discovery of the above edge-darkened and hotspot-less X-shaped
sources might be evidence against   
the backflow model for the creation of wings in at least this class of
sources, the alignment of the   
main radio axis with the optical major axis in these cases as well indicates
that there may be a unified   
model across X-shaped sources with and without hotspots.   
 
A clue is our remarkable finding that the radio images of five of the
eight sources listed above   
clearly show evidence for an inner double---they are double-double radio
sources---and are presumably   
undergoing a renewal of central engine activity with new edge-brightened lobes
aligned with the main outer   
lobes. In this picture the main lobes in all these cases are relics of past
activity, which would likely   
have been of edge-brightened hotspot type if the beams in the past
activity were the same as that at the present. 
We are led to believe that the X-shaped radio sources of the edge-darkened
type and those that lack   
hotspots in their relaxed lobes have relic main lobes, which were created by
powerful beams in the past.    
The relic main lobes, together with the wings, might have been created in
that previous activity phase, with the  
wings a consequence of interaction of backflows with the host galaxy
environment as in the backflow   
model.  If the backflow model is applicable for all X-shaped radio sources with and
without hotspots, this would require that the large-scale   
radio structure in all hotspot-less X-shaped radio sources are relics of past
activity that was in the form of hot-spot radio  
sources.  Additionally, this hypothesis also suggests a conversion from
edge-brightened to edge-darkened   
morphology when the central beams of hot-spot radio galaxies are turned off,
at least in those cases where  
backflows are strong and the beams are directed along the major axis of the
optical host.  

It may be noted that none of the above sources, which are hotspot-less and edge-darkened
X-shaped radio sources, have radio powers significantly below the FR-I/FR-II break, consistent with the hypothesis that they are actually
relics of objects with hotspots.
 
\section{Radio morphology and the radio-optical position-angle offset: the 3CRR sample} 
 
 \citet{cap02} demonstrated that, as projected on the sky, FR-II X-shaped
 radio galaxies tend to have their main radio axis aligned   
closer to the major axis of the host elliptical galaxy.  As we suggest in the
previous section, X-shaped radio   
galaxies that lack hotspots and 
with FR-I type edge-darkened morphology do exist and also follow this trend.
X-shaped radio galaxies are   
a small fraction of extended radio sources associated with AGNs, and it is of
interest to examine whether   
subsets of the parent population, which exhibit other specific morphological
features, also show similar   
preferences in the orientation of their main radio axis relative to the
projected axes of the host,   
and whether relationships exist between morphological parameterizations and
the radio-optical misalignment   
angle. 
 
The results of several previous searches for correlations between the
orientations of the radio axes and   
the host galaxy major or minor axes have been varied. Some early authors, for
example, \citet{sul75},   
\citet{gib75} and \citet{bat80}, all failed to find any correlations. However,
\citet{pal79}   
and \citet{gut80} found a preference for the orientation of the radio axis
along the host minor axis;  
moreover, the tendency for this alignment was stronger for larger  
($>176$~kpc) radio sources. \citet{bir85} ruled out large misalignment angles
($>50\degr$) between the radio   
axes and the minor axes of their hosts.  \citet{san87} examined the
radio-optical axis relations in a large  
sample of sources for which they obtained optical CCD images using the 1.5-m
La Silla and radio images with the Very Large Array (VLA). They report a lack of 
any correlation of the radio axes with the projected optical axes as well as
intrinsic stellar axes.    
More recently, the searches for correlations have been extended  
to examining for differences in correlations amongst subsets of the radio
galaxy population: for example,   
\citet{and99} find a greater preference for FR-II type radio sources to lie
closer to the host minor axis,  
as compared to FR-I type sources, and the correlation was reported to be
stronger for radio sources with   
higher axial ratios.   
 
In this section, we revisit this issue by examining the 3CRR sample for
relationships between radio morphological properties   
and the relative orientation ($\Delta{\rm{PA}}$) between the radio axis and
projected major axis of the host elliptical.  This sample has the advantage of
systematic, precision HST observations with uniform quality, as well  
as exhaustive radio observations available in the literature.  We used the 3CRR
sub-sample \citep{lai83} that   
included only the nearby ($z<0.5$) radio sources (the so-called ATLAS sample
of J.~P. Leahy, A.~H. Bridle   
and R.~G. Strom). These sources have been observed at radio wavelengths with
high sensitivity to extended low-surface brightness  
components and the sample is well suited for examining for preferences in
$\Delta$PA related to radio morphology. 
 
 The radio axes were measured from FITS images available in the 3CRR ATLAS
 website, using detections of core,   
hotspots and jets to define the radio position angle (P.A.). The orientation
of the projected major axes of   
the optical hosts were taken primarily from the tabulations in \citet{dek96}
and \citet{mar99}, who have   
derived the P.A. from HST images.  In all cases we also examined their optical
images by eye and if   
the tabulated value and the apparent P.A. of the optical image appeared
inconsistent, or if a source was not   
included in the tabulations, we (i) used the 1.6-$\mu$m FITS images
downloaded from the HST   
archives, where available, and measured the P.A. of the elliptical isophotes,
and (ii) in certain cases we used the   
FITS images from the digitized sky survey or 2MASS survey repository.  In each
case we also cross-checked against the body of observational   
results cited in the literature for consistency; a most useful reference work
was the HST observations    
reported by \citet{mad06} and the Issac-Newton Telescope (INT) data of
\citet{roc00}.  The P.A. and ellipticity values adopted for each   
galaxy refer  
to the largest scale structure in the optical images; a similar approach was
adopted by \citet{dek96} for their tabulations.  The  
measurements were based on isophotes that were well above noise and mostly on
linear scales of several   
kiloparsec. The P.A. from independent observations and images and using
different data, where available,   
were found to be consistent and usually within $10\degr$. In cases where
isophotal twists confused the issue,   
we adopted determinations based on the outer isophotes of HST archival images.
The basic data used in the analysis presented herein is in Table~1. 
 
In Fig.~1 we show the distribution of $\Delta{\rm PA}$ for the whole sample of
3CRR FR-II sources. The sources are distributed over the range
0--90$\degr$ in their orientations relative to the host major axes  
and appear to show no preference for orientation closer to the host major or
minor axis. This is  
similar to that reported by several previous studies (using ground based data
prior to the HST).

Considering only the 61 3CRR FR-II radio galaxies, the
median linear size   
is 284~kpc and the median P.A. offset of the radio axis from the optical major axis is  
$46\degr$.    
Fig.~2 shows the plot  
of median projected linear size of the 3CRR FR-II sources binned in intervals
of $\Delta{\rm PA}$. The median linear size appears to increase systematically with
radio-optical P.A. Sources with 
radio P.A. within $45\degr$ of the optical major axis have median   
linear size $164 \pm 69$~kpc, whereas the sources with radio P.A. more than $45\degr$  
away from the optical major axis have median linear size of $314 \pm 39$~kpc; the errors 
in the medians have been determined here, as well as hereinafter, usng the Efron Bootstrap method.
We have compared the distributions of the linear sizes of the major
axis sources ($\Delta{\rm PA} > 45\degr$) with the linear sizes of the minor
axis sources ($\Delta{\rm PA} < 45\degr$) using the non-parametric Wilcoxon-Mann-Whitney
test.  The probability that the major axis sources could have a chance distribution in 
linear sizes as small as that observed is 6.5\%, suggesting that the two 
distributions are genuinely different: the population of minor-axis sources have larger linear
sizes as compared to the population of major-axis sources. 
It is interesting 
that when the sample is binned in $22.5\degr$ ranges, the bin containing
sources with radio axes within   
$22.5\degr$ of the minor axis stands out with 4 giant radio galaxies with
linear size exceeding 700~kpc, whereas the remaining three bins do not contain
more than two giant sources each. The Binomial probability of obtaining four or more giants,
of the total of 7 giants in the sample, by chance is 7\% suggesting that our finding of
a preference for giant radio sources to lie close to the host minor axis may be genuine.
Such a minor-axis preference for giant radio galaxies was reported in earlier studies of 
\citet{pal79} and \citet{gut80}.

Interestingly,  
although not large in projected linear size, all the three Relaxed-Double
radio sources in the 3CRR sample  
(3C310, 3C314.1 and 3C386) have radio axes within $22.5\degr$ of the minor
axis; they are all less than about
400~kpc in size.  
 
The sample of 3CRR radio galaxies with FR-II morphological
classification, has 10 giant radio sources with projected linear  
size exceeding 700~kpc (3C33.1, 3C35, 3C46, 3C223, 3C236, 3C274.1, 3C326,
3C457, 4C73.08, DA240).  Of these ten, two have host  
galaxies with circular shapes (3C46 and 3C274.1) and for one source (3C457) we
could not locate useful   
optical images in the literature or repositories.  The remaining seven giant
radio sources, for which the radio-optical P.A.   
may be reliably measured, have a median P.A. offset between the radio axis and
the host minor axis of   
$18\degr$ (the one standard deviation upper limit to this offset is $38\degr$).  
Four out of the seven giant radio sources have their radio axes
within $20\degr$ of the host minor   
axis, and an interesting aspect of the radio morphologies of these large radio
galaxies is the complete   
absence of `winged' structures; we will return to this point later below. The
3CRR FR-II sources, with 
linear sizes below 700~kpc, have a median offset of $43\degr \pm 9\degr$ from the 
major axis. 

It appears that jets in larger radio sources are observed to be preferentially
oriented closer to the minor axis of the hosts;   
additionally, sources close to the minor axes of elliptical galaxies tend to
have larger linear sizes. Radio   
jets may be uniformly distributed in radio-optical P.A. offset; however, it
appears that radio jets that   
propagate closer to the host minor axis result in larger radio sources as
compared to jets that emerge closer   
to the host major axis.  
 
We identified 13 3CRR radio galaxies---including FR-I and FR-II type---that
show prominent wing-like   
distortions in their lobes in the vicinity of the core (3C20, 3C28, 3C61.1,
3C76.1, 3C192, 3C285, 3C288, 3C315,   
3C401, 3C433, 3C438, 4C12.03 and 4C73.08). Eight have wing distortions in both
the lobes whereas five show a   
winged distortion in just one of the lobes. Only one source, 3C28, has mirror
symmetric wings in which both the wing distortions are towards  
the same side of the radio axis. The median radio-optical  
P.A. offset, $\Delta {\rm PA}$, for the 10 sources in this sub-sample 
that have clearly elliptical hosts was found to be $22\degr \pm 7\degr$.  
The median linear size of the sub-sample consisting 
of the 13 winged 3CRR sources is $155 \pm 82$~kpc, which is significantly smaller than the
median linear size of $294 \pm 41$~kpc   
for the entire 3CRR sample. 
The largest source in this sub-sample is the giant radio galaxy 4C73.08, and
the second largest is 4C12.03, both have $\Delta {\rm PA}$ in the range
40--46$\degr$. Of the remaining 11 sources in the sub-sample 8 have
estimates for $\Delta {\rm PA}$, and seven of these have
$\Delta {\rm PA}$ less than $30\degr$.  
 
The indication is that radio galaxies with prominent lobe distortions in the
form of wings are preferentially   
oriented closer to the host major axis, as is the case for X-shaped radio
sources, and such sources are also   
relatively smaller in linear size. The above analysis indicates that winged
distortions are relatively   
uncommon, and perhaps do not form, in giant radio sources and when the jets
are closer to the minor axes.   
 
We may draw particular attention to 3C433, which is one among the thirteen
3CRR sources with prominent wings.    
The source has wings in both lobes, but only one wing is close to the host
galaxy.  While the southern lobe   
extends west close to the location of the host, the extension in the northern
lobe towards east occurs well   
away from the host. The radio axis is close to the major axis of the host:
$\Delta {\rm PA}$ is  
just $20\degr$. Examination of the optical field shows a close neighbour at
the location where the wing   
associated with the northern lobe appears to deflect east. The coincidence
between the bend in radio structure   
and sky location of the companion galaxy is suggestive of a model wherein the
backflow from the northern lobe   
is deflected by a gaseous halo associated with the neighbouring galaxy.  
 
 If sources with prominent wings in their lobes are preferentially oriented
 close to their host major axes, it  
is of interest to examine the morphologies of radio sources that are, instead,
oriented close to their host   
minor axis. We formed a sub-sample of 13 3CRR FR-II radio sources 
(3C35, 3C79, 3C153, 3C219, 3C223, 3C236, 3C244.1, 3C303, 3C310, 3C319, 3C326,
3C390.3, 3C442A) consisting of all sources oriented within   
$\pm 25\degr$ of their host galaxy
minor axis.  The median linear size of these minor-axis sources is $346 \pm 141$~kpc.
Remarkably, only two among these 13 sources (3C310, 3C442A)  
have winged structures associated with their lobes.  Nine sources (3C35, 3C79,
3C219, 3C223, 3C244.1, 3C303, 3C310, 3C319 and 3C390.3) have
continuous bridges---the lobes extend   
all the way to the cores on both sides---and two of the sources (3C153 and 3C442A) have
filled lobe emission on one side. Wing-type distortions are absent in all except
two among these filled lobes.   
This sample of minor-axis radio   
sources includes five giant radio sources. This examination of the sub-sample
of 13 minor-axis sources that have radio axes aligned close  
to their host minor axes reveals that most are devoid of wings---including any
scaled down manifestations of   
wing distortions or mini-wings---in their lobes.   
 
We may also examine the morphologies of radio galaxies selected to have radio
axes close to their host major   
axes.  Sixteen FR-II-type radio galaxies (3C67, 3C200, 3C28, 3C436, 3C173.1,
3C401, 3C268.3, 3C299, 3C61.1, 3C33.1, 3C433, DA240, 3C388, 3C42, 3C300
and 3C123) were identified in the 3CRR sample,
which have radio axes within $\pm 25\degr$ of their host major axes.  
The median linear size of this major-axis source sample is just $149 \pm 63$~kpc.  Half
of the sources have  
mini-wings, wings or plume structures associated with their lobes and five
(3C28, 3C401, 3C61.1, 3C433 and 3C123) of
these eight display prominent wings. However, it may be noted that some of
these major-axis sources have no wings (3C436, 3C173.1 and 3C33.1) and some
have large emissions gaps between their cores and lobes (3C299, 3C268.3 and 3C67).
There is only one giant radio galaxy, DA240,  
in this restricted sample; although DA240 does not have lobe distortions such
as wings or mini-wings, both   
its lobes have unusually small axial ratios (ratio of the lobe extent along
the radio axis, from core to the end,   
to the transverse width of the lobe), and are individually nearly circular in morphology. 
 
 To conclude this section, there appears to be two classes of radio galaxies:
 major-axis and minor-axis   
sources that have differing radio structures and different median linear
sizes. Most major-axis sources have relatively smaller extents and often
exhibit off-axis lobe distortions; minor-axis sources appear to have relatively  
larger linear sizes and lack off-axis wing distortions in their lobes.  
  
\section{Radio morphology and the radio-optical position-angle offset: 
the Giant Radio Galaxy sample}
 
 In this section we explore the radio-optical P.A. offsets in the 1-Jy sample
 of giant radio galaxies of   
\citet{sch00}, restricting to the sub-sample of redshift $z < 0.15$ objects
for which optical images that   
have sufficient quality for the reliable determination of P.A.s exist.  The
sub-sample consists of 17   
giant radio galaxies (GRGs) with projected sizes exceeding 700~kpc; five of
these are members of the 3CRR   
sample.  The methodology adopted for estimating the radio and optical P.A.s
was the same as for the 3CRR sample. Three sources (B0050+402, B0658+490  
and B2147+816) had poor optical images and were rejected; of the remainder,
one host galaxy (B1309+412) had a shape close to circular and was also omitted for the
analyses that required $\Delta$PAs.
 
 We derive a median offset of $11\degr \pm 3 \degr$ between the radio axis and host minor axis
 for the reduced 
sample of 13 GRGs with $z<0.15$ and size $>700$~kpc; as many as 11 of the 13
have radio axis within $30\degr$   
of the host minor axis. The median radio-optical P.A. offset for the 3CRR  
giant radio galaxies (with size $>700$~kpc) is consistent, within errors,  with the value obtained 
for this GRG sample; additionally, it may be noted that
the omission of the five 3CRR giants from this GRG sample does  
not change the computed median.   
 
 The analysis of the 3CRR sample indicated that relatively few GRGs manifested
 large wings. To explore this   
issue for a larger sample, we combined complete samples of GRGs with good
radio images: the 11 3CRR giants,   
24 giants from the 1-Jy sample of \citet{sch00} (excluding B0050+402 for which
a good radio image could not   
be found in the literature), 7 giants from the sample of \citet{sub96}
(excluding B2356$-$611 that has a   
linear size $<700$~kpc) and the SUMSS sample of 16 giant radio galaxies from
\citet{sar05}.  It may be noted   
here that the 11 3CRR sources and the 7 sources in \citet{sub96} were
well-imaged; additionally, the sources   
in \citet{sch00} and \citet{sar05} are from surveys, with high
surface-brightness-sensitivity, at 327~MHz   
and 843~MHz respectively.  Avoiding repetitions, we have 52 GRGs in all with
linear size $>700$~kpc.  We   
examined their radio images for off-axis lobe distortions close to the centre:
off-axis plumes with extent   
$> 50\%$ of the associated lobes are referred to herein as `wings'; those $<
50\%$ of their associated lobes are   
referred to as `mini-wings'.    
 
 A large fraction ($87\%$) of the giant radio galaxies in this compilation do
 not have wings. Only 3 (4C73.08, B0114$-$476 and B0511$-$305) of the   
52 have extensions resembling wings, and in all of these sources the wings are
in only one of the pair of   
lobes. One source, 3C46, has a pair of mini-wings and another three (3C274.1,
B0211$-$471 and B1545$-$321) have
mini-wings on only one side.    
Clearly wings of the kinds seen in X-shaped radio galaxies are not present.
If we restrict to only the   
18 well-imaged sources---the 3CRR giants and those in the compilation of
\citet{sub96}---we find that 11   
do not have wings, 3 have wings (only in one lobe), 3 have mini-wings (only in
one lobe) and one has a pair   
of mini-wings, confirming the relative paucity of wings in GRGs that was inferred for
the larger compilation.  
 
To examine the relationship between the radio axes and host optical axes for a 
larger sample of GRGs, we used the combined sample, but restricted to sources 
for which relatively good optical data is available: (a) the 11 3CRR
GRG hosts for 10 of which HST  
images are available, (b) an additional 9 sources in the \citet{sch00} sample
of nearby $z<0.15$ 1-Jy giants (excluding the 5 common
to 3CRR), for which  
sufficiently good DSS or 2MASS images were available, and (c) 4 of the 7 GRGs 
from the \citet{sub96} sample   
for which Anglo-Australian Telescope (AAT) imaging observations or SuperCOSMOS
digitized images with sufficient   
quality were available. Our measurements were checked against images and notes
from various publications in   
the literature. In all there were 24 sources: Table~2 lists the members from the
$z<0.15$ 1-Jy \citet{sch00} sample (including the 5 common to 3CRR) and the 4
from \citet{sub96} sample; the six members that are part of the 3CRR nearby
sample are in Table~1.  Five GRGs 
(3C46, 3C274.1, B0511$-$305, B1545$-$321 and B1309+412) were associated with hosts with 
apparently circular  
images and one (3C457) had only a poor optical image and, consequently, host
P.A.s could not be reliably estimated for these. 

Fig.~3 shows the histogram of the $\Delta$PA distribution for
the 18 GRGs with non-circular hosts and for which optical-axis P.A.s
could be determined; a clear 
preference for minor axis orientation is seen. 
Among these 18 GRGs, 12 have radio axes within $30\degr$ of the host minor
axes whereas 6 have radio axes more than $30\degr$ from the host minor axes.
The chance probability that 12 or more of the 18 giants have radio axes 
within $30\degr$ of the host minor axis is as low as 0.4\% if they were
uniformly distributed.  
The median P.A. offset between the radio axis and host  
minor axis for the 18 GRGs, for which optical host P.A.s could be reliably
measured, is $11\degr \pm 9 \degr$.  This median   
offset is similar to that obtained for the 3CRR giants, and separately for the
13 \citet{sch00} giants.  
 
 We may compare the radio morphologies of two groups of giant radio
 galaxies---those with radio axes   
within 30$\degr$ of the host minor axis and those with larger offsets.  All
the 12 GRGs with radio axis   
close to the host minor axis lack central extensions: both wings and
mini-wings. Several of these GRGs show   
emission gaps; what is noteworthy is that in every case where the lobe
emission extends all the way to the   
central core no wings or mini-wings are apparent in the radio images 
(e.g. 3C35, 3C223 and B0309+411).  In the
6 GRGs with radio axis more   
than 30$\degr$ offset from the host minor axis, winged lobes are manifest in 4.
4C73.08 and B0114$-$476 have wings in one lobe each and B0211$-$479 and DA340 have
a mini-wing each.  
 
 To summarize this section on GRGs, examination of the radio morphologies and
 comparison with P.A.s of   
their optical host galaxies shows that a significant fraction of GRG radio
axes are close to the optical   
minor axes and that these minor-axis GRGs are relatively deficient in lobe
distortions in the form of wings   
and mini-wings.  The few GRGs with radio axes oriented away from the minor
axes of their hosts are relatively   
abundant in extended wings and mini-wings.

\section{Radio morphology and the radio-optical position-angle offset: 
X-shaped radio sources}

 In this section the distribution of radio-optical P.A. offsets in X-shaped
 radio sources are examined and   
relationships between the source morphology and radio-optical P.A. offset are
studied.  A total of 31 X-shaped   
radio sources have been compiled from the literature.  Selection criteria
adopted were: (a) presence of   
off-axis lobe distortions at the end of the lobe closer to the core, which
exceed 50\% of the length of the associated lobe,  
(b) the distortion ought to be asymmetric with respect to the main source
axis, and (c) the off-axis    
distortion must be present in at least one of the lobes in the pair.  These
criteria exclude the numerous   
sources with mini-wings.  To the list of known X-shaped radio sources compiled
by \citet{che07} (see his Table~1)  
we added sources from the 3CRR sample and from   
the Abell clusters imaged by \citet{owe97}, which  
were recognized as satisfying the criteria for X-shaped structure. There are
two quasars in the compilation.    
The total radio powers span a wide range from $3\times10^{24}$ to
$10^{27}$~W~Hz$^{-1}$.  Our sample is more than   
three times larger than the sample used by \citet{cap02} in their study of the
distribution in radio-optical P.A. offset.  
 
 X-shaped radio galaxies are believed to be exclusively associated with FR-II
 type radio sources \citep{cap02}.   
However, in our compilation of 31 sources we identify as many as 8 that are
either edge-darkened or lack   
hotspots at the lobe ends despite being edge-brightened.  All have been
described above in Section 2   
along with a discussion on the implications for the backflow model for X-shaped radio sources. 
 
 In Table~3 we list parameters for our sample of X-shaped radio sources. For deriving the  
P.A.s of the main radio axis we adopted the same method we used---and
described above---for the 3CRR sources.   
For obtaining P.A.s for the optical hosts: in the case of 3CRR sources we used
the values derived above; for   
the remaining sources the optical P.A.s were measured on the DSS or
SuperCOSMOS images at a surface brightness   
level at least three times above rms noise. We expect measurement errors to be
less than $10\degr$. For   
measurement of P.A. of the radio wings we mostly used the P.A. of the central
ridge line of individual wings.   
The wings in a pair of lobes are usually parallel, although the wings are
often not collinear; in a few cases   
(3C315 and 3C403) we adopted the mean P.A. for the two wings and in cases
where only one wing was well   
defined its P.A. was used.

 Fig.~4, which shows the distribution in P.A. of the main radio axes and wings
 for our sample, reveals that   
the main radio axes in X-shaped radio sources are all oriented within
$50\degr$ of their host major axis with   
median offset of $25\degr \pm 6 \degr$. The Fig.~4 also shows the different distributions
in the P.A. offsets for the   
wings compared to the main radio axes: whereas the main radio axes are fairly
uniformly distributed over   
offsets in the range 0--50$\degr$ of the host major axes, and only two sources
have main radio axes within   
$10\degr$ of the host major axes, the wings are preferentially close to the
host minor axes and as many as   
8 have wings within $10\degr$ of their host minor axis. The probability
that the different distributions for the orientations of the main sources
and the wings occurs by chance is $<0.01$\% based on the Wilcoxon-Mann-Whitney test.
If we exclude the two quasars and sources with faint or unidentified
or circular host galaxies, we have 10  
sources whose wings are within $10\degr$ of the minor axis of their hosts
and 9 sources with wings making larger angles. 
These distributions imply that although the main   
radio axis might be somewhat offset from the host major axis, the wings tend
to align along the host minor   
axis. These confirm the work of \citet{cap02}. 
  
Fig.~5 shows a plot of the ratio of projected extents of wing to main source versus
source size. The two quasars (4C01.30 and J2347+0852) have large wings relative to  
their main radio axes, as would be expected because their main radio axes
would be foreshortened by projection effects.  
Considering the radio galaxies in the sample, the median projected
wing-to-main-lobe ratio shows no evidence for a trend with projected linear
size: the median value of this ratio is $0.91 \pm 0.05$.
The distribution over this ratio appears to have a long tail towards large
values of the wing-to-main-lobe ratio.  Fig.~5 suggests that sources with relatively larger
projected linear size, including giant radio sources, are lacking in examples
of sources with large wing-to-main-lobe ratio. Such a trend is consistent with
a picture wherein asymmetry in galaxy environment results in larger sources
being minor-axis sources with poorer backflows and off-axis distortions in their lobes.
However, because of the small size of the sample it is difficult to correct for
projection effects, which may be partly responsible for this trend.

The scatter plot in Fig.~6, showing the distribution in radio power versus the
P.A. of the wings relative to the major axis of the host, is indicative of a
weak link between the orientation of the wings with   
respect to the host major axis and the source total radio power (correlation
coefficient between the wing position angle and the logarithm of the total radio 
power is 0.55). The X-shaped
radio sources in which the wings are well away from the host  
minor axis are the ones with relatively lower total radio power, and the more
powerful X-shaped radio sources tend to  
have wings along the host minor axes.  Although their main radio axes might
make angles of as much as   
$50\degr$ to the host major axes, as is the case for 4C12.03, 3C403 and
B2356$-$611, these powerful sources continue to have minor  
axis wings. 
 
In Fig.~7 we show the distribution of the ellipticity of the host galaxies of
X-shaped sources compared to that for 3CRR FR~II hosts.  
The hosts of the X-shaped
sources appear in the plot to have a distribution to greater ellipticities and 
the median ellipticity for the hosts of X-shaped sources is $0.28 \pm 0.05$, 
which is somewhat greater than the median ellipticity 
($0.21 \pm 0.03$) of the hosts of 3CRR FRII radio sources. 
The indications here are in agreement with an earlier suggestion \citep{cap02} of relatively higher
ellipticity for the hosts of X-shaped sources.
However, it may be noted here that this finding is not statistically
significant since the 
Wilcoxon-Mann-Whitney test suggests a 26\% probability that the observed 
distribution to higher ellipticities is a chance occurance.
As pointed out by \citet{spr07}, X-shaped radio galaxy hosts need not be highly   
elliptical relative to the parent population of hosts of radio galaxies. In
our sample there are 6 sources   
with apparently circular host galaxies.  As galaxy ellipticity is known  
to vary with radius we need a more quantitative measure for this parameter,
and high quality images of a sample   
of hosts, to examine for correlations with more confidence.

\section{Central lobe extensions as mini wings?} 
 
 We have also examined the radio-optical axis offsets in radio sources with
 mini-wings: inversion-symmetric central distortions that result in extensions
 of the lobes transverse to the main radio axis, which   
have lengths less than half the length of their respective main lobes.  Our
sample of mini-wing sources consists of 11 members of the 3CRR sample: 3C19,
3C33, 3C42, 3C46, 3C234, 3C274.1, 3C284, 3C319, 3C341, 3C300 and 3C381.  Because the
P.A.s of the mini wings are ill   
defined, we examined only the radio-optical P.A. offsets of main lobes with
respect to the optical axes. 
 
 Five of the host galaxies appear circular, in 2 sources (3C42 and 3C300) the
 radio axis is within $25\degr$   
of the host major  
axis, and in 4 (3C33, 3C284, 3C319 and 3C381) the radio axes make angles exceeding 
$50\degr$ relative to
their host major axes.  We   
examined the four cases of large offsets individually. In two sources, 3C284
and 3C381, the P.A. offset between   
the radio axes and extended emission-line gas is about $35\degr$ and $30\degr$ respectively 
\citep{mcc95}. It is remarkable that the lobe deflections resulting in mini wings appear to
occur near the boundary of the   
emission line gas distribution and the directions of apparent bending are
consistent with a deflection of   
backflow at the boundary.  In the case of 3C319 the radio axis is close
to the minor axis ($\Delta$PA   
= $78\degr$): in this case the mini-wing is observed in only one lobe and here
we identify a galaxy to the   
south of the host which appears to have deflected the backflow
\citep{kee95}. In the fourth case, 3C33, extended emission line gas at P.A. of
about 20$\degr$ with respect to the southern lobe appears to be influencing the
backflow and deflecting it East, whereas the northern lobe deflects East somewhat
away from the core \citep{bau88}.  A member of this sample,   
3C341, is noted to  
have a circular galaxy image: emission-line gas is observed to be extended over
32~kpc and oriented nearly   
North-South and the relative angle between the radio axis and the gas major
axis is about $50\degr$ \citep{mcc95}. 
To summarize, in these mini-winged sources the P.A. of the emission-line gas
distribution appears to replace the role of the host galaxy in the shaping of the backflows.

\section{Summary and discussion} 
 
The problem for the backflow model posed by the discovery of a class of
X-shaped radio sources   
with edge-darkened lobes and lobes that lack hotspots is alleviated by  
the discovery of double-double morphologies in a relatively large fraction of
them, which suggests that these X-shaped radio sources appear so only
because their central engines have recently ceased feeding their outer lobes.
Excluding the two with circular hosts, all the six sources we have identified as
belonging to this class have their main radio axes closer to the   
host major axis and wings that are closer to the host minor axis---properties
associated with FR-II X-shaped sources.  Moreover, the lobe-wing pairs are
observed to be on the same side of the major axis, as required by
the backflow model for the formation of X-structures. Therefore, it   
appears that whether of FR-I type or FR-II type the backflow model is able to
explain the formation of wings in these sources. The inference that X-shaped
sources that appear to have an FR-I structure were 
FR-II type hot-spot radio sources in the past active phase, with strong backflows
that interacted with the host galaxy, leads to the suggestion that the
edge-brightened FR-II lobes
transformed to edge-darkened FR-I morphology with time following the cessation of
jets, at least for these major-axis sources with strong backflows. 

For the 3CRR sample of nearby radio sources we have shown that there is no
preferred orientation of the  
radio axis: radio galaxies emerge at all angles with respect to the host major
axis. This lack of a   
preferred orientation for a general population of radio galaxies was also the
conclusion of several   
previous investigations (\citet{san87} and references therein). 
However, examining the properties of large-scale radio structures in the
nearby 3CRR powerful radio galaxies, clear differences are seen 
depending on the direction of propagation of their jets relative to
their host axes. Sources  
that are oriented along a direction closer to the host major axis---the
major-axis sources---do not grow large
on average and tend to be 
associated with winged lobe structures. In contrast, minor-axis sources 
have larger projected linear sizes on the average and rarely develop
off-axis distortions in their lobes.  
 
Separately, for a sample of giant radio galaxies, we have shown that there is
a strong preference for the radio axis to be   
along the host minor axis: giant radio sources are predominantly minor-axis
sources. While a small fraction do form along directions closer to the host 
major axes these are found   
to have prominent off-axis lobe structures. Once again, similar to the 3CRR
sample, major-axis giant radio sources are   
found to be associated with winged structures. 
 
For our sample of 31 X-shaped sources, which is larger than the
sample examined by \citet{cap02}, we find a striking preference
for the main radio axes to be in the quadrants of the host major axes, and for
the wings to be in the quadrants containing the host minor axes; this is
consistent with and an improvement on the significance of the earlier work.

\subsection{Asymmetry in galaxy environment: influence on forward and backflows}

Several previous studies demonstrated a link between environment and radio structures: 
\citet{mcc91} presented the correlated asymmetries between radio source extents and extended 
emission-line gas distribution, and \citet{sar86} and \citet{sub08} for
correlated asymmetries   
between radio source extents and ambient large-scale environments of giant
radio galaxies. That the ambient  
medium affects radio morphologies is also demonstrated by examples like 3C433 and
NGC~326, in which the   
lobes on the side of the close galaxy neighbour develop off-axis deviations
correspondingly earlier, and examples  
like 3C341 and 3C28, in which the extended emission-line region appears to shape and
redirect plasma flows although influence of an overdensity of hot, X-ray
emitting gas, particularly in 3C28, may have a greater influence (e.g., \citet{har07}, 
\citet{eva08}). In all of these examples it is thermal gas 
concentrations that are encountered by radio jets or
bulk synchrotron plasma flows, resulting in a hindrance to the forward flow
or a deflection of backflows.  Our findings of a clear connection between the
morphological properties of powerful radio galaxies and their orientation
relative to their host---whether the radio source is a minor- or major-axis
source---is suggestive of an influence of asymmetry in the host galaxy
environment on the radio source morphology.  The previous case studies and
findings of circumstantial evidence for interactions between radio synchrotron
plasma and thermal gas suggest that our work presented herein may be
interpreted as arising due to differences in the ambient density and gradients
thereof along the host galaxy major axis compared to that along the host
galaxy minor axis.  These differences may differentially affect the forward
propagation of jets in the major and minor axis directions and the strength of backflows;
additionally, the off-axis deflection of backflows would themselves differ
between minor- and major-axis sources owing to the different strengths in the
backflows and pressure gradients. 

It may be noted here that although `mini-wing' type lobe distortions have been
associated with thermal gas concentrations in the form of  
extended emission line regions, the more spectacular `wings' in X-shaped radio
sources appear to be a consequence of lobe interactions with the more extensive
thermal halos of galaxies. 
 
Since double radio sources are observed
to have similar axial ratios independent of linear size \citep{sub96}, it follows that
backflow---at least of the relativistic 
electrons---is essential if the bridge is to remain visible despite the
inevitable expansion losses.  Studies have indicated that backflowing plasma
from hotspots in powerful radio 
galaxies might attain quite high velocities: perhaps as much as several tenth's of 
the velocity of light \citep{ale87}.  The conditions encountered by the
backflowing plasma in the  
vicinity of the host galaxy in the two cases of major- and minor-axis jet
propagation are different.  Assuming that the gaseous environment of the host
galaxy has the ellipticity and P.A. of the stellar component, the gas density
and pressure gradients would be greater along the minor as compared to the
major axis.

\subsection{Clues from the `wings' in double radio sources}

In most winged radio sources and the X-shaped radio sources 
an association in the form of a connectedness in structure is often observed
between the wing pair and the lobe pair: each of the wings are associated with
particular lobes and in many it has been observed that the associated wing and
lobe are on the same side of the host major axis.  The connection between lobe
and associated wing has sometimes also been observed in a continuity in
B-field.  These argue in favor of models in which wings are fed by backflows
from lobes.

It is remarkable that wings are observed exclusively in sources that have
continuous or filled main lobes: there is no case in which the main lobes are
docked and a pair or wings are observed as a symmetric structure extending
across the core, disconnected from the main lobes, and forming a secondary
pair of lobes orthogonal to the main lobes. 
Sources with continuous bridges are expected to be those with
strong backflows and, therefore, this finding underlines the importance
of backflows, and the channeling of backflows into wing structures, 
for the existence of visible wings.

Mechanisms suggested for the backflow deflection have been either a buoyant
lifting of the plasma away from  
the dense thermal galactic halo gas or a force provided by an overpressured
cocoon of synchrotron plasma  
that then seeks the direction of the steepest pressure gradient along the
minor axis. The wings are often
observed to have rather collimated morphologies with magnetic field mostly
along the linear extent.   
The integrity of the deflected backflow may not be maintained if it were to
just expand freely; therefore,  
while buoyancy may be playing a role, a driving force is required. An
overpressured cocoon as   
suggested by \citet{cap02} is a possibility, although rather than originating
only at the ends of the lobes it may be that the overpressure develops
naturally as the backflow  
stalls upon reaching the denser regions of the galactic halo. Collimation in
the main lobes is thought of as arising due to
greater advance speeds at the ends of the lobes compared to the lateral
expansion speeds of the sides of the lobes.  Collimation in the wing
structures may arise for a similar reason: within the backflow model and in an
asymmetric environment the wings advance most rapidly along the direction of
steepest pressure and density gradient.  Since the steepest gradient is along the
minor axis of the host galaxy, then this might also be the preferred direction
for the orientation of the wings, consistent with the finding in Fig.~4.

As noted earlier in
Section~5, the median ratio of wing to lobe lengths is 0.91 for X-shaped
sources, implying that there are as many X-shaped sources with this ratio
exceeding 0.91 as those with this ratio in
the range 0.5--0.91.  The suggestion that 
large wing-to-main-lobe ratios are seen only in relatively smaller radio
sources, and that this ratio decreases with increasing source
size, is consistent with evolution of radio sources in an
anisotropic environment, which 
manifests in a retardation in the growth of major-axis sources, enhancement in
their backflows, and formation of wings of length dependent on the strength of
the backflows and the retardation of the forward advance of the jets.

Within the backflow scenario it is possible for extreme asymmetry in the
environment to result in wings advancing faster than the main lobe, resulting
in wings larger than the main lobe extent. This would require the main lobes
to be directed along the direction of greater density, and the backflow
creating wings along the direction of lower density.  Assuming that the heads
at the ends of the main lobes as well as the advancing ends of the wings are
ram pressure limited in their speeds by the external density, the ratio of the
external densities encountered by the main lobes and wings would have to
exceed the ratio of the pressures in the heads and wings. Such an explanation
for X-shaped radio sources in which the wings are larger than the main source
extent weakens the argument that the existence of such structures requires axis
flips for their origin.

Concerning the wings in X-shaped radio sources, we may list three findings
that argue in favor of a pre-existing channel along the minor axes 
or a preferred plane normal to the major axes as a conduit for the backflowing
material: (1) the collimation
observed in many wings of X-shaped radio sources, (2)
the larger extents observed for many wings compared to the main lobes, and (3)
the finding, as shown in Fig.~4, that wings tend to align close to the minor axes
of the hosts even though the main radio axes have a wider spread in 
distribution about the major axes. Our discussions above suggest that these
findings may indeed have explanations within the backflow model, indicating
that a purely environmental explanation is possible.

\subsection{The black-hole merger model}
 
First, it may be noted that 'dead' radio galaxies are known to be
notoriously rare \citep{blu00}. Once the jets cease feeding the lobes,   
they are expected to suffer severe adiabatic expansion  
losses and disappear over a relatively short timescale compared to source
lifetimes. In this context, it is difficult to understand the wings of
X-shaped radio sources as relics of past activity when the renewed activity
following a jet-reorientation has evolved to a linear size exceeding that of
the wing, and particularly in the case of giant X-shaped radio sources.

Second, the morphologies of wings in X-shaped radio sources bear little
resemblance to the known relic    
double radio sources like IC2476 \citep{cor87} and SGRS~J1911$-$7048 \citep{sar05},
both of which have an overall edge-brightened and bounded appearance.
It is perhaps more appropriate to compare the wings with relics that have
been `dead' for timescales similar to that in which axis flips might occur:
this has been estimated to be about $10^{7}$~yr \citep{mer02}. The outer
relic lobes of double-double
radio galaxies, which are believed to be restarting, have been estimated to
have been `dead' for such timescales \citep{sch00,sar02,saf08}; however,
none of the wings in X-shaped radio sources resemble these relics. 
Unlike the range of morphologies observed for the general population of radio
galaxies, all winged structures appear to be continuous  
and well filled with no emission gaps towards their central regions. It is
unlikely that axis flips would occur only in a narrow range of source
morphologies.

The indications are that it is unlikely that X-shaped radio sources represent
a pair of relic lobes and another independent 
pair of lobes created following an axis flip. 
In this context, it is of interest to consider the model proposed by
\citet{lal07}, wherein the twin lobes are thought of as separate radio sources
associated with a close pair of AGNs. A compelling argument against such a model is
that the morphologies of the main lobes and wings of X-shaped sources
have specific relationships and are not a
random sampling of the morphologies observed in double radio sources. The main 
lobes are almost always edge-brightened and have bridges extending to
the core. The wings do not have hotspots and they too are always extended all
the way to the core. It is also difficult to account for the
lobe-wing pairing seen in several X-shaped sources and the minor axis preference for
the weaker and edge-darkened lobe pair in such a model.

In an alternate model suggested by \citet{lea84},
X-shaped sources might be a result of an axis flip following which backflow from
the new lobes rejuvenates the relic lobe maintaining its visibility.  In this
picture a pre-existing channel along the minor axes 
as a conduit for the backflowing material might be naturally present as a
relic cocoon, if the earlier activity prior to the black-hole merger and
associated axis flip were aligned with the host minor axis.  
 
Consistent with the finding that there is no preferred alignment between the
radio axis and host optical axis in a general population of radio sources, we
might expect all possible orientations of the black hole spin axis 
prior to a merger and associated flip. Within the black-hole merger model,
flips from minor-axis to major-axis result in the new source experiencing
enhanced backflow and, consequently a tendency to manifest as an X-shaped
radio source if the backflow illuminates the relic lobes closer to the minor
axis. The question then arises as to the outcome of axis flips from an initial
major-axis state to a minor-axis orientation.

In this context, of relevance is the recent discovery of non-thermal X-ray
emission in the high redshift radio galaxy, 3C294, on either side of the core,
and at a P.A. offset of about $50\degr$ from the main radio axis; the X-ray
lobes are not visible in the radio \citep{erl06}.  Axis flips from a major- to
minor-axis source would be expected to create new lobes along the minor axis,
with poor backflow, less likelihood of forming an X-shaped radio source, but
would result in a minor-axis double radio source with a relic pair of
radio-invisible lobes closer to the major axis that might be visible in its
inverse-Compton scattering of cosmic microwave background radiation.  


\subsection{The role of black-hole axis re-orientation}

A final point is regarding our finding of a tendency for giant radio sources to
preferentially be oriented closer to the host minor axis. In the unified model
presented above seeking a common explanation for the formation of giants,
X-shaped radio sources and off-axis lobe distortions, anisotropy in the galaxy
environment is offered as the cause. We comment on another aspect here.

If the minor axis in elliptical galaxies represents the axis of the large  
scale galaxy potential, it is the direction to which an offset black-hole axis may be  
re-oriented, as a result of an opposing torque to the torque that the black
hole exerts on the immediate  
accretion disk surrounding it in the attempt to position it in its equatorial
plane \citep{nat98}.   
This re-alignment timescale has been estimated to be $\sim10^{6}$~yr, for
Eddington accretion rates, but could be larger for lower
accretion rates.  

The median linear size of radio sources 
are observed to be larger for orientations closer to the host minor axis (Fig.~2),
which might be interpreted as alignment of the radio axis with the host minor
axis with age.  If this realignment is owing to the above \citet{nat98} model,
we may expect that re-alignment timescales are comparable to lifetimes of
giant radio galaxies, and that accretion rates are significantly below
Eddington.  However, since the distribution of radio axes of sources is fairly
uniform (Fig.~1), implying that sources with relatively smaller sizes---presumably
younger---are major-axis sources, the onset of 
radio activity is coeval with an event that offsets black hole axes away from the
stable minor-axis direction.

\section{Conclusions} 
 
We have presented a new class of X-shaped radio sources, which lack hotspots
and are sometimes edge-darkened. Many of this class are observed to 
have double-double radio source morphologies. 

We have explored relations between radio
source morphologies and the relative angle between the radio axis and host
optical galaxy axes.  The study is based entirely on axis
orientations as projected on the sky.   The well known 3CRR sample
of nearby radio galaxies has been  
used given the high quality radio and optical data available for this
sample. Radio-optical morphology relations have been also examined for
representative samples of giant radio galaxies and, separately, X-shaped radio
sources. The main conclusions of our study are: 
 
(1) Wing formation in both edge-brightened as well as edge-darkened X-shaped radio sources may
be understood as arising from redirection of backflow into wings. 
Most edge-darkened X-shaped radio sources are presently undergoing renewed nuclear activity
that is observed in the form of inner double structures: we propose that
the wings in edge-darkened X-shaped radio sources represent backflows from a
past activity episode.  
 
(2) FR-II type edge-brightened lobes, formed by powerful beams in hot-spot
radio sources, may evolve into edge-darkened FR-I type morphology following
cessation of jet activity.   
 
(3) Although double radio sources may have axes randomly distributed with
respect to the host galaxy axes, the size to which a   
radio source grows and its radio morphology are influenced by the relative orientation.
Minor-axis sources tend to have larger linear sizes compared to major-axis sources.
Major-axis sources have a greater propensity for off-axis lobe distortions in
the form of winged structures as compared to minor-axis sources. 
 
(4) Giant radio galaxies are preferentially oriented along their host minor axes.

(5) Winged radio galaxies are predominantly major-axis sources. 
 
(6) X-shaped radio sources have radio axes within $50\degr$ of their host
major axes. The wings in X-shaped radio sources are within 40$\degr$ of their
host minor axes.  Although the main radio axes in these sources are fairly
uniformly distributed over the range, the wings have a strong preference to be
within about 10$\degr$ of the host minor axes. 
  
(7) In X-shaped radio sources with higher radio power, the wings have a
relatively higher propensity to be oriented closer to the host minor axis. 

We are led to the conclusion that the morphology of powerful double radio
sources depends on whether the source is of major- or minor-axis type.  The
reason appears to be an anisotropy in the environment, which is related to the
ellipticity of the host galaxy.  The anisotropy appears to influence the forward advance of the
beams, the strength of the backflow in the cocoon, and the formation of wings
and off-axis distortions in the lobes. 
The interaction of backflows in a major-axis radio source with an anisotropic
environment associated with 
the host---perhaps an elliptic halo---appears necessary for the
formation of X-shaped radio sources; this may indeed be a sufficient
condition.  
 
There is observational evidence for binary nuclei and axis flips in AGNs.
Axis flips may not be a necessary or dominant mechanism for the
formation of X-shaped radio sources. Nevertheless, the
study presented here suggests that even if axis flips are
the cause for the formation of X-shaped radio sources, backflows appear to be important
in directing material from the new lobe into the relic lobes.

The lack of any preferred radio-optical P.A. for the general population of
radio sources suggests that the initial orientation of an active radio source prior
to a black-hole merger might not have any preferred direction.  Assuming this
is so, an interesting outcome is that if X-shaped radio
sources are created by axis flips in the black-hole merger model, an equal
number of minor-axis radio sources ought to exist, representing sources in which the
axis has flipped from a major- to minor-axis orientation, in which the relic
lobe along the major axis might be observable in inverse-Compton X-rays and
not in radio emission.  However, if black holes are re-oriented to align with
the host minor axis over time, black-hole mergers and associated axis flips
might result in post-merger radio sources that avoid the minor axis.

\acknowledgments

This publication makes use of data products from the Two Micron All Sky
Survey, which is a joint project of the University of Massachusetts and the
IPAC/California Institute of Technology, funded by NASA and the NSF.  This
research has made use of data obtained from the SuperCOSMOS Science Archive,
prepared and hosted by the Wide Field Astronomy Unit, IoA, University of
Edinburgh, which is funded by the UK PPARC.  The
Digitized Sky Surveys were produced at the STScI under US Government grant NAG
W-2166.

\clearpage
\begin{deluxetable}{llllllll} 
\rotate
\tablewidth{6.7 in}
\tablecolumns{8}
\tablecaption{The 3CRR sample} 
\tablehead
{ \colhead{Name} & \colhead{Redshift} & \colhead{FR class} & \colhead{PA-Radio} &
\colhead{PA-Optical} & \colhead{$\Delta$PA} & \colhead{ellipticity} & \colhead{Reference} }
\startdata 
3C16	&	0.405	&	II	&	28.8	&	-22*	&	50.8	&	0.28*	&	deV 2000	\\ 
3C19	&	0.482	&	II	&	29.3	&	***	&		&	0	&	deV 2000	\\ 
3C20	&	0.174	&	II	&	103.13	&	130	&	26.87	&	0.28	&	deK 1996	\\ 
3C28	&	0.1971	&	II	&	149.7	&	155	&	5.3	&	0.39	&	deK 1996	\\ 
3C31	&	0.0173	&	I	&	162*	&	144	&	18	&	0.12	&	Mar 1999	\\ 
3C33.1	&	0.181	&	II	&	44.5	&	63	&	18.5	&	0.26	&	deK 1996	\\ 
3C33	&	0.0595	&	II	&	19.2	&	148*	&	51.2	&	0.26*	&	Bau 1988	\\ 
3C35	&	0.067	&	II	&	9.6	&	111	&	78.6	&	0.26	&	Mar 1999	\\ 
3C42	&	0.395	&	II	&	131.6	&	153	&	21.4	&	0.47	&	deK 1996	\\ 
3C46	&	0.4373	&	II	&	65	&	***	&		&	0	&	deV 2000	\\ 
3C61.1	&	0.186	&	II	&	2.1	&	165	&	17.1	&	0.18	&	deK 1996	\\ 
3C66B	&	0.0215	&	I	&	45*	&	135	&	90	&	0.05	&	Mar 1999	\\ 
3C67	&	0.3102	&	II	&	175.3	&	175	&	0.3	&	0.4	&	deK 1996	\\ 
3C76.1	&	0.0324	&	I	&	139	&	140*	&	1	&	0.12*	&	MAST	\\ 
3C79	&	0.2559	&	II	&	101	&	25*	&	76	&	0.16*	&	MAST	\\ 
3C83.1	&	0.0179	&	I	&	100	&	166	&	66	&	0.22	&	Mar 1999	\\ 
3C98	&	0.0306	&	II	&	21	&	55	&	34	&	0.13	&	Mar 1999	\\ 
3C109	&	0.3056	&	II	&	151	&	***	&		&	0	&	Bau 1988	\\ 
3C123	&	0.2177	&	II	&	114.9	&	90	&	24.9	&	0.32	&	deK 1996	\\ 
3C132	&	0.214	&	II	&	130.8	&	7	&	56.2	&	0.25	&	deK 1996	\\ 
3C153	&	0.2769	&	II	&	48	&	125*	&	77	&	0.21*	&	MAST	\\ 
3C171	&	0.2384	&	II*	&	100.9	&	***	&		&	0	&	MAST	\\ 
3C173.1	&	0.292	&	II	&	18.5	&	8	&	10.5	&	0.23	&	deK 1996	\\ 
3C184.1	&	0.1182	&	II	&	161.8	&	40	&	58.2	&	0.18	&	deK 1996	\\ 
3C192	&	0.0598	&	II	&	123.3	&	***	&		&	0	&	Mad 2006	\\ 
3C200	&	0.458	&	II	&	155.3	&	150	&	5.3	&	0.36	&	deK 1996	\\ 
3C219	&	0.1744	&	II	&	39.6	&	145	&	74.6	&	0.08	&	deK 1996	\\ 
3C223	&	0.1368	&	II	&	163.2	&	75*	&	88.2	&	0.1*	&	Mad 2006	\\ 
3C234	&	0.1848	&	II	&	62	&	***	&		&	0	&	McL 1999	\\ 
3C236	&	0.0989	&	II	&	122.8	&	44	&	78.8	&	0.44	&	Mar 1999	\\ 
3C244.1	&	0.428	&	II	&	168	&	76	&	88	&	0.2	&	deK 1996	\\ 
3C268.3	&	0.371	&	II	&	160.6	&	144	&	16.6	&	0.46	&	deK 1996	\\ 
3C272.1	&	0.004	&	I	&	0	&	126	&	54	&	0.17	&	Mar 1999	\\ 
3C274	&	0.004	&	I	&	114	&	0.3	&	66.3	&	0.03	&	Mar 1999	\\ 
3C274.1	&	0.422	&	II	&	75.5	&	***	&		&	0	&	deK 1996	\\ 
3C284	&	0.2394	&	II	&	101.4	&	42*	&	59.4	&	0.19*	&	MAST	\\ 
3C285	&	0.0794	&	II	&	82.8	&	129*	&	46.2	&	0.55*	&	MAST	\\ 
3C288	&	0.246	&	II*	&	141	&	***	&		&	0	&	deK 1996	\\ 
3C293	&	0.04503	&	II	&	128.1	&	66	&	62.1	&	0.54	&	Mar 1999	\\ 
3C295	&	0.4599	&	II	&	144.3	&	***	&		&	***	&	Bau 1988	\\ 
3C296	&	0.0237	&	I	&	36.5	&	148	&	68.5	&	0.21	&	Mar 1999	\\ 
3C299	&	0.367	&	II	&	63.6	&	47	&	16.6	&	***	&	deK 1996	\\ 
3C300	&	0.27	&	II	&	126.7	&	104	&	22.7	&	0.47	&	Roc 2000	\\ 
3C303	&	0.141	&	II	&	98.6	&	0.0*	&	81.4	&	0.1*	&	deK 1996	\\ 
3C305	&	0.041	&	II*	&	44.2	&	76	&	31.8	&	0.37	&	Mar 1999	\\ 
3C310	&	0.054	&	II	&	160.9*	&	86	&	74.9	&	0.18	&	Mar 1999	\\ 
3C314.1	&	0.1197	&	I	&	144*	&	75	&	69	&	0.31	&	deK 1996	\\ 
3C315	&	0.1083	&	I	&	8.4	&	33	&	24.6	&	0.46	&	deK 1996	\\ 
3C319	&	0.192	&	II	&	43.5	&	146	&	77.5	&	0.17	&	Roc 2000	\\ 
3C321	&	0.096	&	II	&	134.7	&	***	&		&	***	&	Mar 1999	\\ 
3C326	&	0.0885	&	II	&	78.3	&	150*	&	71.7	&	0.5*	&	Roc 2000	\\ 
3C341	&	0.448	&	II	&	53.4	&	***	&		&	0	&	Roc 2000	\\ 
3C346	&	0.161	&	II*	&	72.6	&	122	&	49.4	&	0.23	&	deK 1996	\\ 
3C349	&	0.205	&	II	&	146	&	14	&	48	&	0.31	&	deK 1996	\\ 
3C381	&	0.1605	&	II	&	3.5	&	58*	&	54.5	&	0.18*	&	MAST	\\ 
3C382	&	0.0578	&	II	&	51	&	85	&	34	&	0.22	&	Mar 1999	\\ 
3C386	&	0.017	&	I	&	14*	&	102	&	88	&	0.13	&	Mar 1999	\\ 
3C388	&	0.0908	&	II	&	61.3	&	40	&	21.3	&	0.13	&	Mar 1999	\\ 
3C390.3	&	0.0569	&	II	&	149	&	82	&	67	&	0.16	&	Mar 1999	\\ 
3C401	&	0.201	&	II	&	16	&	0.3	&	15.7	&	0.123	&	Roc 2000	\\ 
3C433	&	0.1016	&	II*	&	164.5	&	145*	&	19.5	&	0.47*	&	deK 1996, MAST	\\ 
3C436	&	0.2145	&	II	&	176.4	&	3	&	6.6	&	0.31	&	deK 1996	\\ 
3C438	&	0.29	&	II	&	128.4	&	***	&		&	***	&	MAST	\\ 
3C442A	&	0.0263	&	II	&	64*	&	131	&	67	&	0.13	&	Mar 1999	\\ 
3C449	&	0.0171	&	I	&	10*	&	1	&	9	&	0.2	&	Mar 1999	\\ 
3C452	&	0.0811	&	II	&	75	&	101	&	26	&	0.32	&	Mar 1999	\\ 
3C457	&	0.428	&	II	&	40	&	***	&		&	***	&	***	\\ 
3C465	&	0.0313	&	I	&	123.2	&	30	&	86.8	&	0.22	&	Mar 1999	\\ 
4C12.03	&	0.157	&	II	&	22.63	&	160	&	42.63	&	0.32	&	2MASS	\\ 
4C14.11	&	0.206	&	II	&	140	&	***	&		&	***	&	***	\\ 
4C14.27	&	0.392	&	II	&	107.5	&	***	&		&	***	&	***	\\ 
4C73.08	&	0.0581	&	II	&	70	&	25	&	45	&	0.26	&	2MASS	\\ 
DA240	&	0.035	&	II	&	61.23	&	40	&	21.23	&	0.14	&	2MASS	\\ 
NGC6251	&	0.024	&	I	&	111	&	30	&	81	&	0.1	&	2MASS	\\ 
\enddata 
\tablenotetext{*} {For optical: quoted value either differs from the value in
  deK96 or is not present in
  deK96. Used MAST FITs data for   
3C76.1, 3C79, 3C153, 3C284, 3C285 and 3C381. For others, values were
measurements based on data in references cited. For radio: P.A.  
of jets were measured from higher resolution images or from central regions
in the case of relaxed doubles.}  
\tablenotetext{***} {Indeterminate either because the object is confused or
  there is lack of relevant data in literature} 
\tablerefs{Bau1988---Baum et al. 1988; deK1996---de Koff et al. 1996;
  Mar1999---Martel et al. 1999;   
Roc2000---Roche \& Eales 2000; deV2000---de Vries et al. 2000; Mad2006---Madrid et al. 2006;  
McL1999---McLure et al. 1999; MAST---Multimission Archives at STScI} 
 
 
\end{deluxetable} 
 
\clearpage
 
\begin{deluxetable}{lllllll}
\rotate 
\tablewidth{0pc}
\tablecolumns{7}
\tablecaption{Radio-optical data for giant radio galaxies from \citet{sch00} (z<0.15) and 
\citet{sub96} with good optical images} 
\tablehead{ 
\colhead{IAU-Name}&\colhead{Redshift}&\colhead{FR
  class}&\colhead{PA-Radio}&\colhead{PA-Optical}&  
\colhead{$\Delta$PA}&\colhead{Reference}} 
 
\startdata 
B0055+300 (NGC315) 	&	0.0167	&	I	&	135	&	55	&	80	&	2MASS	\\ 
B0109+492 (3C35)	&	0.067	&	II	&	9.6	&	111	&	78.6	&	Mar 1999	\\ 
B0157+405 (4C40.08)	&	0.0827	&	II?	&	109	&	66	&	43	&	DSS	\\ 
B0309+411 	        &	0.134	&	II	&	130	&	47	&	83	&	DSS	\\ 
B0648+733	        &	0.1145	&	II	&	52	&	-40	&	88	&	2MASS	\\ 
B0745+560 (DA240)	&	0.035	&	II	&	61.23	&	40	&	21.23	&	2MASS	\\ 
B0945+734 (4C73.08)	&	0.0581	&	II	&	70	&	25	&	45	&	2MASS	\\ 
B1003+351 (3C236)	&	0.0989	&	II	&	122.8	&	44	&	78.8	&	Mar 1999	\\ 
B1209+745 (4CT74.17.1)	&	0.107	&	I	&	154	&	53	&	79	&	2MASS	\\ 
B1309+412	        &	0.1103	&	II	&	6	&	***	&		&	2MASS	\\ 
B1312+698 (DA340)	&	0.106	&	II	&	127	&	11	&	64	&	DSS	\\ 
B1426+295	        &	0.087	&	II	&	45	&	-35	&	80	&	DSS	\\ 
B1626+518	        &	0.0547	&	II	&	35	&	-65	&	80	&	2MASS	\\ 
B1637+826 (NGC6251)	&	0.024	&	I 	&	111	&	30	&	81	&	2MASS	\\ 
	&		&		&		&		&		&		\\ 
B0114-476	        &	0.146	&	II	&	172	&	7	&	15	&	SuperCOSMOS	\\ 
B0211-479	        &	0.2195	&	II	&	173	&	145	&	28	&	SuperCOSMOS	\\ 
B0511-305	        &	0.0583	&	II	&		&	***	&		&	SuperCOSMOS	\\ 
B1545-321	        &	0.1085	&	II	&		&	***	&		&	SuperCOSMOS	\\ 
 
\enddata 
 
\tablenotetext{***}{Indeterminate because optical image appears circular} 
 
 
 
\end{deluxetable} 

\clearpage

\begin{deluxetable}{llllllllllll} 
\tabletypesize{\footnotesize}
\rotate
\tablewidth{0pc}
\tablecolumns{12}
\tablecaption{The X-shaped source sample}  
\tablehead{ 
\colhead{Name}&\colhead{Redshift}&\colhead{FR}&\colhead{LLS}&\colhead{P$_{1.4}$}& 
\colhead{PA-Radio}&\colhead{PA-Wing}&\colhead{PA-Optical}&\colhead{elipticity}&\colhead{W/M}& 
\colhead{Ref}&\colhead{Ref} \\ 
\colhead{}&\colhead{}&\colhead{class}&\colhead{(kpc)}&\colhead{(W/Hz)}& 
\colhead{(deg.)}&\colhead{(deg.)}&\colhead{(deg.)}&\colhead{}&\colhead{}&
\colhead{Optical}& \colhead{Radio} 
} 
 
\startdata 
4C12.03	&	0.156	&	II	&	623	&	26.11	&	22.6	&	65	&	160	&	0.32	&	1.06	&	2MASS	&	Lea 1991	\\ 
B0037+209 	&	0.0622	&	II	&	189	&	24.05	&	119	&	40	&	85	&	0.456	&	0.57	&	DSS	&	Owe 1997	\\ 
(Abell 75)	&		&		&		&		&		&		&		&		&		&Led 1995		&		\\ 
3C20	&	0.174	&	II	&	155.2	&	26.34	&	103.13	&	17	&	130	&	0.28	&	0.72	&	deK 1996	&	3CRR atlas	\\ 
NGC326	&	0.047	&	II	&	112.6	&	24.95	&	135	&	55	&	***	&	0	&	2.13	&	Cap 2000	&	Mur 2001	\\ 
J0116-473	&	0.146	&	II	&	1896	&	26.21	&	170	&	70	&	2	&	0.21	&	0.76	&	DSS	&	Sar 2002	\\ 
3C52	&	0.2854	&	II	&	329	&	26.99	&	22	&	114	&	20	&	0.36	&	0.88	&	deK 1996	&	Lea 1984	\\ 
3C63	&	0.175	&	II	&	84.7	&	26.47	&	34	&	130	&	79	&	0.26	&	1.53	&	deK 1996	&	Har 1998	\\ 
3C76.1	&	0.0324	&	I 	&	128	&	24.7	&	139	&	57	&	140*	&	0.12*	&	0.87	&	MAST	&	3CRR atlas	\\ 
3C136.1	&	0.064	&	II	&	543.1	&	25.47	&	107	&	0	&	117	&	0.39	&	0.97	&	Mar 1999	&	Lea 1984	\\ 
3C192	&	0.0598	&	II	&	229	&	25.63	&	123.3	&	35	&	***	&	0	&	0.66	&	Mad 2006	&	Bau 1988	\\ 
3C197.1 	&	0.1303	&	II	&	47	&	25.79	&	0	&	71	&	***	&	0	&	1	&	deK 1996	&	Owe 1997	\\ 
(Abell 646)	&		&		&		&		&		&		&		&		&		&		&		\\ 
4C32.25	&	0.0512	&	II	&	338	&	25.06	&	64	&	0	&	84	&	0.15	&	0.83	&	Ulr 1996	&	Mac 1994	\\ 
3C223.1	&	0.1074	&	II	&	163.4	&	25.77	&	15	&	130	&	40	&	0.45	&	1.4	&	deK 1996	&	Bla 1992	\\ 
4C48.29	&	0.052	&	II	&	459	&	25.03	&	170	&	105	&	***	&	0	&	1	&	DSS	&	Jag 1987	\\ 
J1101+167 	&	0.0677	&	II	&	270	&	24.84	&	114	&	20	&	64	&	0.291	&	1.42	&	DSS	&	Owe 1997	\\ 
(Abell 1145)	&		&		&		&		&		&		&		&		&		&Led 1995		&		\\ 
4C01.30	&	0.1324	&	II	&	116	&	25.49	&	100	&	24	&	85	&	0.3	&	1.97	&	Wan 2003	&	Wan 2003	\\ 
B1142+157 	&	0.067	&	II	&	45	&	24.48	&	59	&	153	&	67	&	0.329	&	0.87	&	DSS	&	Owe 1997	\\ 
(Abell 1371)	&		&		&		&		&		&		&		&		&		&Led 1995		&		\\ 
4C04.40 	&		&	II	&		&		&	35	&	136	&		&		&	1.12	&	---	&	Jun 2000	\\ 
(B1203+043)	&		&		&		&		&		&		&		&		&		&		&		\\ 
J1210+719 	&	0.1226	&	I	&	110	&	24.99	&	162	&	74	&	125	&	0.476	&	0.92	&	DSS	&	Owe 1997	\\ 
(Abell 1484)	&		&		&		&		&		&		&		&		&		&Led 1995		&		\\ 
J1357+4807	&	0.383	&	II	&	57.5	&	26.16	&	18	&	65	&		&		&	1.28	&	---	&	Leh 2001	\\ 
3C315	&	0.1083	&	I	&	394	&	26.1	&	8.4	&	123	&	33	&	0.46	&	0.93	&	deK 1996	&	Lea 1984	\\ 
J1534+103 	&	0.1333	&	I	&	524	&	25.13	&	171	&	72	&	***	&	0	&	0.57	&	DSS	&	Owe 1997	\\ 
(Abell 2091)	&		&		&		&		&		&		&		&		&		&		&		\\ 
3C379.1	&	0.256	&	II	&	340	&	26.57	&	161	&	45	&	***	&	0	&	0.63	&	deK 1996	&	Spa 1985	\\ 
3C401	&	0.201	&	II	&	78.4	&	26.5	&	16	&	109	&	0.3	&	0.123	&	0.78	&	Roc 2000	&	3CRR atlas	\\ 
3C403	&	0.059	&	II	&	129	&	25.68	&	79	&	135	&	39	&	0.25	&	2	&	Mar 1999	&	Bla 1992	\\ 
B2014-55	&	0.06	&	I	&	1513	&	25.18	&	155	&	72	&	10	&	0.4	&	0.57	&	DSS	&	Jon 1992	\\ 
3C433	&	0.1016	&	II*	&	127	&	26.48	&	164.5	&	60	&	145*	&	0.47*	&	0.88	&	deK 1996	&	3CRR atlas	\\ 
	&		&		&		&		&		&		&		&		&		&MAST		&		\\ 
3C438	&	0.29	&	II	&	100	&	27.2	&	128.4	&	40	&	***	&	***	&	0.68	&	MAST	&	3CRR atlas	\\ 
J2157+0037	&	0.3907	&	II	&	427.5	&	26.11	&	69	&	151	&		&		&	1.22	&	---	&	Zak 2004	\\ 
J2347+0852	&	0.292	&	II	&	314.4	&	25.62	&	135	&	52	&	148	&	0.15	&	1.38	&	DSS	&	Lan 2006	\\ 
B2356-611	&	0.0963	&	II	&	736	&	26.68	&	133	&	90	&	3	&	0.37	&	0.91	&	DSS	&	Sub 1996	\\ 
\enddata 
 
\tablenotetext{***}{indeterminate or optical image appears circular} 
\tablerefs{For Optical: Cap 2000---Capetti et al. 2000; deK1996---de Koff et al. 1996;  
Led 1995---Ledlow \& Owen 1995; Mad 2006---Madrid et al. 2006; Mar
1999---Martel et al. 1999;    
MAST---Multimission Archives of STScI; Roc 2000---Roche \& Eales 2000; Ulr
1996---Ulrich \& Roennback 1996;   
Wan 2003---Wang, Zhou \& Dong 2003;  
For Radio: Bau 1988---Baum et al. 1988; Bla 1992---Black et al. 1992; Jag
1987---Jaegers 1987;   
Jon 1992---Jones \& McAdam 1992; Jun 2000---Junor et al. 2000; Lan
2006---Landt, Perlman \& Padovani 2006;   
Lea 1984---Leahy \& Williams 1984; Lea 1991---Leahy \& Perley 1991; Leh
2001---Lehar et al. 2001;   
Mac 1994---Mack et al. 1994; Mur 2001---Murgia et al. 2001; Owe 1997---Owen \&
Ledlow 1997;  
Sar 2002---Saripalli, Subrahmanyan \& Udaya Shankar 2002; Spa 1985---Spangler
\& Sakurai 1985;  
Sub 1996---Subrahmanyan, Saripalli \& Hunstead 1996; Zak 2004---Zakamska et al. 2004} 
 
 
 
\end{deluxetable} 

\begin{figure}
\includegraphics[scale=0.6,angle=0]{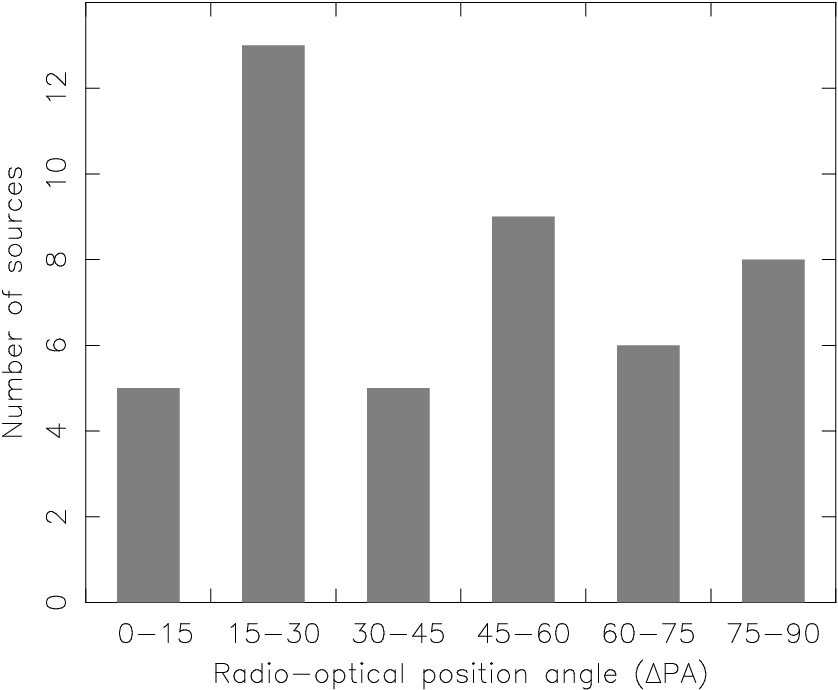}
\caption{ Histogram of $\Delta$PA for the 3CRR FR-II sources.}
\end{figure}

\begin{figure}
\includegraphics[scale=0.6,angle=0]{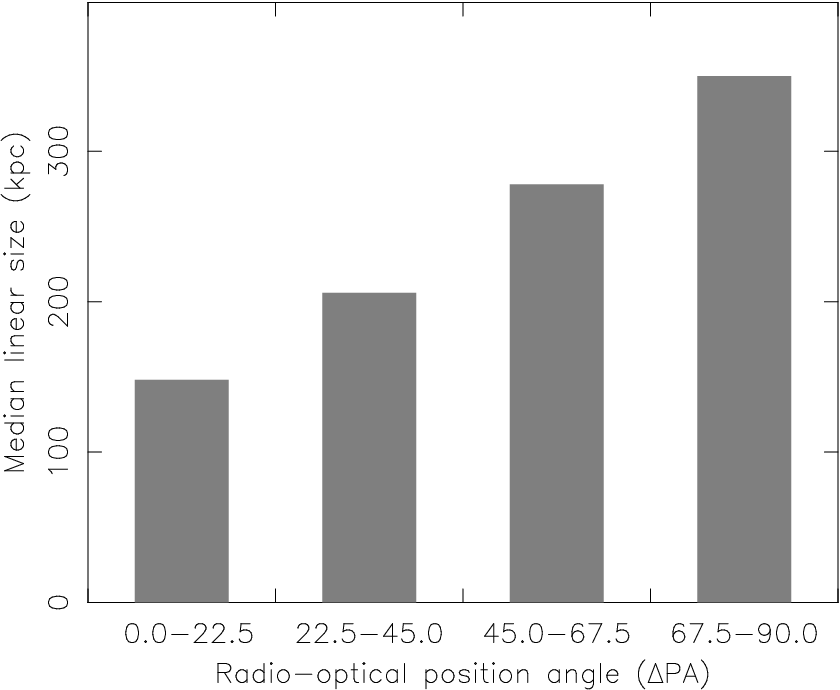}
\caption{ Bar graph of the median projected linear size, in kpc, for the 3CRR
  FR-II radio sources binned in intervals of $\Delta$PA.}
\end{figure}

\begin{figure}
\includegraphics[scale=0.6,angle=-90]{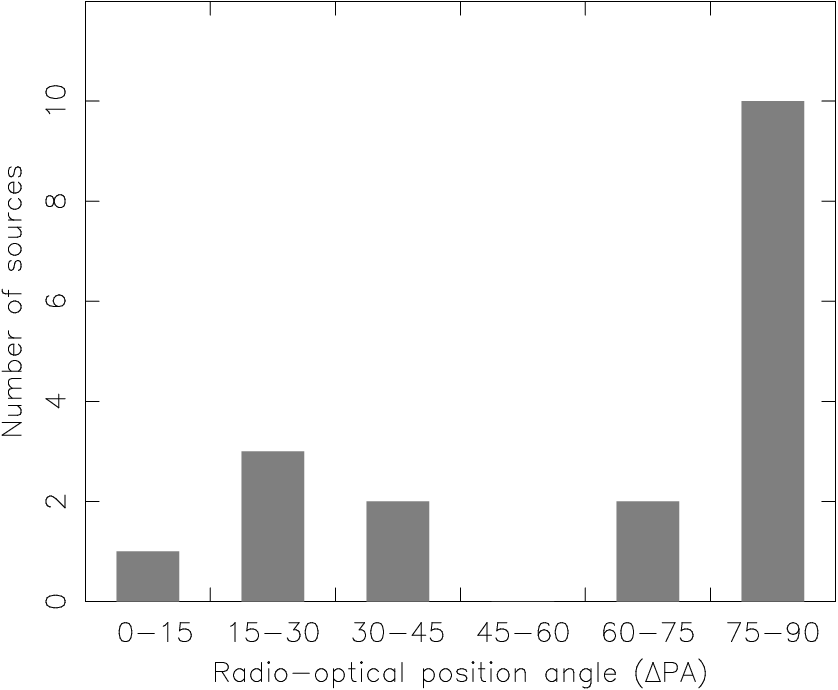}
\caption{ Histogram of $\Delta$PA for the giant radio sources.}
\end{figure} 

\begin{figure}
\includegraphics[scale=0.6,angle=-90]{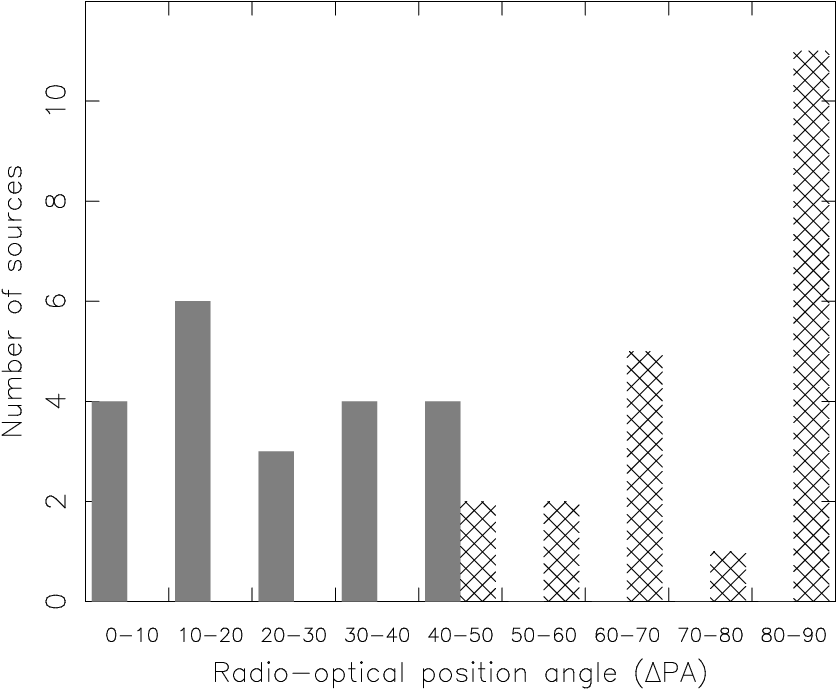}
\caption{ Histogram of $\Delta$PA for the X-shaped radio sources: the
  $\Delta$PA distribution for the main source axis is shown using shaded bars and
  the $\Delta$PA for the wings is shown as cross-hatched bars.}
\end{figure}

\begin{figure}
\includegraphics[scale=0.6,angle=0]{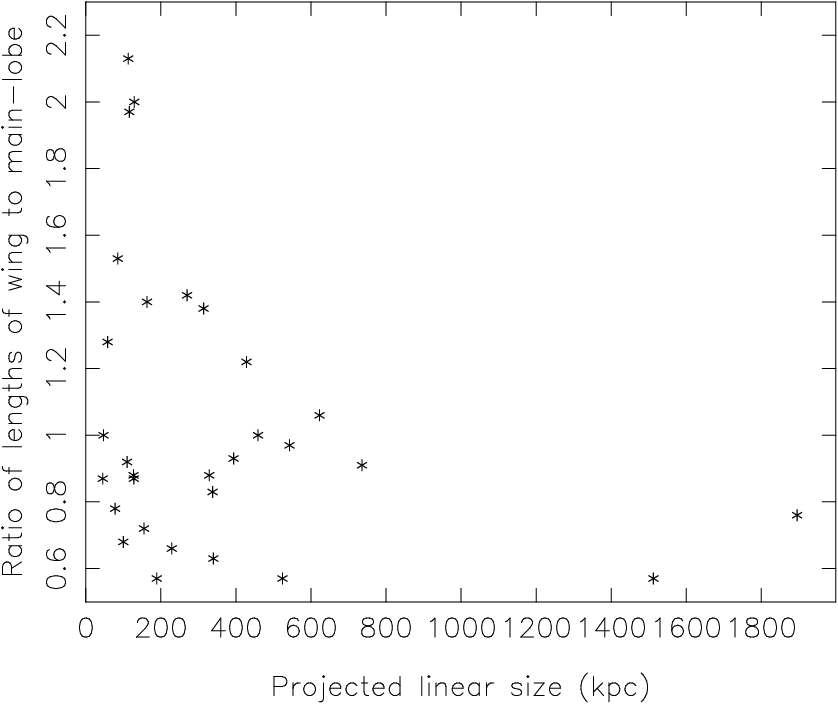}
\caption{ Plot of the ratio of projected lengths of wings to main lobes, for
  the sample of X-shaped radio sources, versus the projected linear size.
}
\end{figure}

\begin{figure}
\includegraphics[scale=0.6,angle=-90]{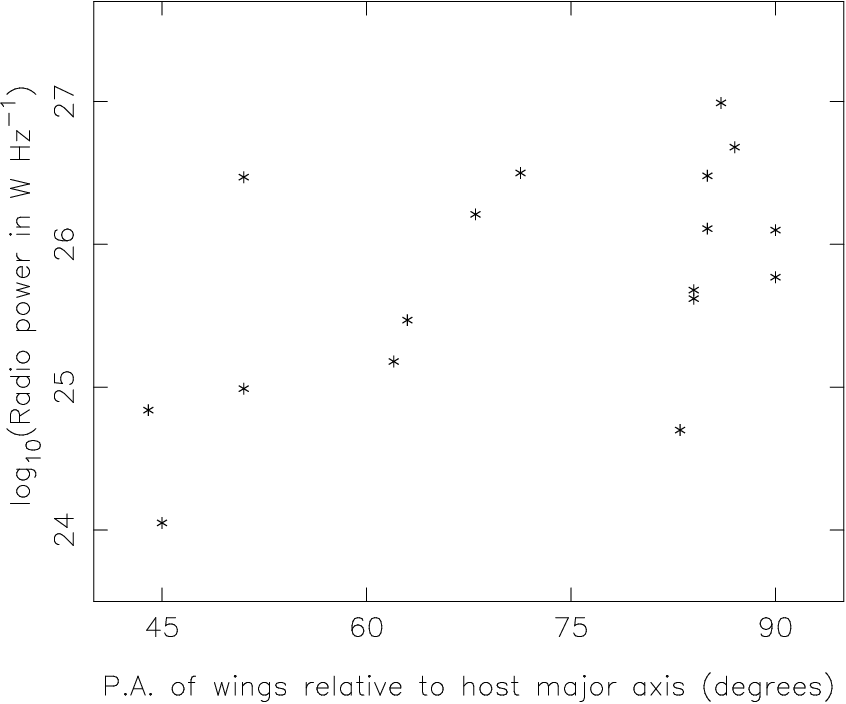}
\caption{ Plot of the radio power of the X-shaped radio sources versus the
  position angle of the wings relative to the major axis of the host.
}
\end{figure}

\begin{figure}
\includegraphics[scale=0.6,angle=-90]{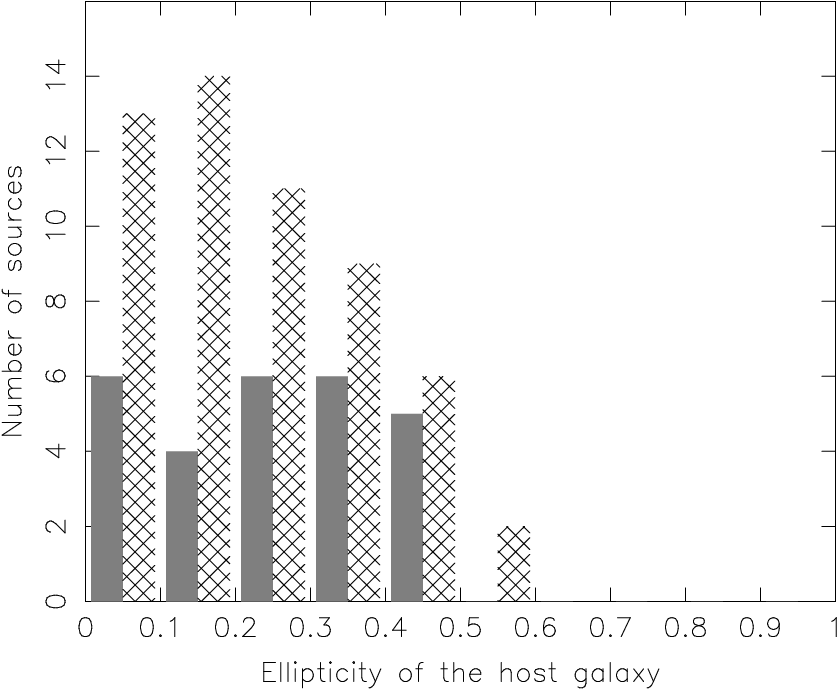}
\caption{ Histogram of ellipticities of the host galaxies for the X-shaped
  radio sources compared with that for hosts of 3CRR FR-II sources. The bars
  for the X-shaped radio sources is shown using shaded bars and
  that for the 3CRR FR-II hosts is shown as cross-hatched bars.}
\end{figure}

\end{document}